\documentclass[twocolumn,showpacs,epsfig,pra]{revtex4}

\usepackage{graphicx}
\usepackage{amssymb}
\usepackage{color}

\newcommand{\figsize}{0.43}
\newcommand{\figsizesmall}{0.4}

\begin{document}

\draft
\title{Ray model and ray-wave correspondence in coupled optical microdisks}
\author{Jung-Wan Ryu}
\email{jwryu@pks.mpg.de}
\author{Martina Hentschel}
\affiliation{Max-Planck-Institut f\"ur Physik komplexer Systeme, D-01187 Dresden, Germany}
\date{\today}

\begin{abstract}
We introduce a ray model for coupled optical microdisks, in which we select coupling-efficient rays among the splitting rays. We investigate the resulting phase-space structure and report island structures arising from the ray-coupling between the two microdisks. We find the microdisks's refractive index to influence the phase-space structure and calculate the stability and decay rates of the islands. Turning to ray-wave correspondence, we find many resonances to be directly related to the presence of these islands.
We study the relation between the (ray-picture originating) island structures and the (wave-picture originating) spectral properties of resonances, especially the leakiness of the resonances which is represented as the imaginary part of the complex wave vector.
\end{abstract}
\maketitle
\narrowtext

\section{Introduction}
With the advances in material science and nano-fabrication techniques, coupled optical microdisks have recently attracted much interest especially in the context of device applications such as photonic molecules and coupled-resonator optical waveguide (CROW).  The terminology `photonic molecule' is introduced as an optical analog to chemical molecules. Their molecule-like structure arises from pairs of interacting optical microdisks (that, in the analogy, represent the individual atoms) that are coupled by narrow channels \cite{Bay98,Muk99,Har03,Nak05,Ish05,Bor06_1}. CROW is the chain version of an optical molecule, and this chain of coupled optical microdisks can be used as an optical delay line through weak coupling between localized optical high-$Q$ cavities \cite{Yar99,Ste98}. In addition, coupled optical microdisks have also been studied for theoretical prospects and experimental realizations in the contexts of optical mode coupling and producing directional light emission \cite{Ryu06,Bor07,Ben08,Sha08,Ryu09}.

The ray picture and the associated ray dynamics have been investigated extensively in optical microcavities since asymmetric resonant cavities, whispering-gallery resonators with smooth deformations of a circular boundary shape, were introduced in order to break the rotational invariance of disk resonators and obtain directed emissions from optical microcavities \cite{Cha96,Noc97}. In those slightly deformed microcavities, light is predominantly emitted from the boundary points of highest curvature and the direction is tangential to the boundary. From a ray dynamical viewpoint, this is a result of tunneling of the rays supporting the whispering gallery mode (WGM) confined in the Kolmogorov-Arnold-Moser (KAM) tori through the lowest dynamical barrier. As the microcavity is strongly deformed, the ray dynamics becomes more chaotic and the direction of light emission is determined by the unstable manifold structure \cite{Sch04,Lee05,Shi06,Lee207,Wie08,Tan07}. There are many recent successes in understanding and designing microcavities based on a ray dynamical analysis that explained the existence of high-$Q$ modes with directional emission characteristics.

However, in coupled optical microcavities, the identification of a ray dynamical model and consequently the establishment of ray-wave correspondence are not simple because of the intrinsic refractive ray splitting. Quantum chaos in systems with ray splitting has been studied by ray models with stochastic selection rule \cite{Cou92,Blu96,Blu96_2,Sir97,Koh98,Pue03,Pue04} and for special resonator geometries such as annular cavities \cite{Hen02}. In this paper, we introduce the ray model for coupled optical microdisks, in which we select coupling-efficient rays among the splitting rays by imposing a deterministic selection rule and study ray-wave correspondence based on this ray model. In Sec. II we explicate our ray model in coupled dielectric disks and show how island structures in phase space originate from the coupling between two disks. We also obtain the stabilities and decay rates of principal periodic orbits as well as the island structures. Regular modes localized on islands are presented in Sec. III and we study the relation between classical structures
and resonance modes systematically. Finally, we summarize the results in Sec. IV.

\section{Ray model with deterministic selection rule for coupled dielectric disks}

The ray model of a dielectric disk is very simple. If the incident angle of a ray is larger than the critical angle for total internal reflection, the ray totally reflects and, due to the rotational symmetry and the related conservation of angular momentum, circulates inside the disk forever with the same incident angle. If the incident angle of a ray is smaller than the critical angle, the ray reflects and transmits partly -- the ray is refractively splitted. The loss of rays to the outside implies that the overall intensity inside the disk will be reduced according to (a generalized) Fresnel's law \cite{Hen202}, but apart from this, all rays inside the disk will circulate keeping the initial angle as the incident angle for each reflection and no additional features arise. The ray model of a two-disk billiard consisting of two closed disks is also very easy to understand: If a ray is inside one of the disks, it will again circulate inside this disk with the same incident angle. If a ray approaches the disks from outside, it will escape after one scattering at the (hard-wall) disks. Note that there exists one unstable periodic orbit in the system, namely the shortest path between the two disks.

\begin{figure}
\begin{center}
\includegraphics[width=\figsize\textwidth]{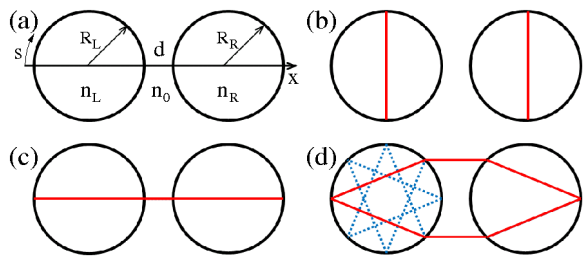}
\caption{(color online) (a) Coupled dielectric disks. Simple periodic orbits (red lines) in coupled dielectric disks. (b) Vertical bouncing ball type periodic orbit. (c) Horizontal bouncing ball type periodic orbit. (d) Hexagonal-shaped periodic orbit. The blue dotted lines represent the associated quasiperiodic (or periodic) trajectories in a single disk.}
\label{fig1}
\end{center}
\end{figure}

However, the ray model of two coupled dielectric (i.e., partially open) disks with radii $R_L$ and $R_R$, refractive indices $n_L$ and $n_R$, and a separation $d$ [cf.~Fig.~\ref{fig1} (a)] is not simple because the ray which emits from one disk can, in principle, enter the other disk. At the recurrence of the ray into the first disk, the incident angle will be varied, and the dynamical properties of a ray model for coupled dielectric disks become plentiful, as will become visible in a more complicated phase-space structure.
An additional complication arises from the fact that the ray splitting leads to an exponential multiplication of rays, namely to $2^n$ trajectories after $n$ bounces, that are, in practice, impossible to follow individually. Therefore, a selection of rays has to be made, based on a selection rule that picks those rays that determine the system dynamics and properties. There are many possibilities for imposing a selection rule; we have chosen a deterministic selection rule that emphasizes the coupling between the two disks: 
Among the two splitted rays, we select the one that is more effective for the coupling between the two disks - if one of the splitted rays (e.g. the transmitted part of a ray approaching the disk boundary from inside one of the disks) will reach the other disk, we choose this one. If this ray would, however, escape the system, we choose the ray that remains inside (enters into) one of the disks. Totally reflected rays do not exhibit ray splitting and remain in the respective disk forever.

A more precise formulation of our deterministic selection rule reads as follows:
If $|p|$, modulus of the sine of the incident angle, of an initial ray inside the left disk is larger than the critical $p_c = n_0/n_L \equiv 1/n$ for total internal reflection, the ray circulates inside the left disk and does not reach the right disk. If the initial ray  is located in the open (refractive) region, i.e. $-p_c<p<p_c$, the ray reflects and transmits partly according to Fresnel's equation \cite{Haw95,Hen202}. The reflected ray remains inside the left disk but the transmitted ray either escapes from the system or approaches the right disk. We select the reflected ray if the transmitted ray emits from the system but the transmitted ray if it reaches the right disk. The ray that hits the right disk also reflects and transmits partly and here we always discard the reflected ray and keep the transmitted ray that always enters into the second disk. We repeat this selective process whenever the rays meet the dielectric boundary. Note that, consequently, none of the kept light rays ever escapes the system. Contrary to the previous studies on ray splitting model with a stochastic selection rule \cite{Cou92,Blu96,Blu96_2,Sir97,Koh98,Pue03,Pue04}, we introduce here a ray model with a deterministic selection rule (RMDS) and apply it throughout this paper.

We first consider periodic orbits in coupled dielectric disks in the framework of the RMDS ray model. In comparison to the single disk, coupling between two disks due to openness produces many new periodic orbits. Figure~\ref{fig1} (b)-(d) show three types of simple periodic orbits in coupled dielectric disks. The vertical bouncing ball type periodic orbits of Fig.~\ref{fig1} (b) are the same as those of a single dielectric disk, except for a rotational symmetry breaking resulting from the coupling. The horizontal bouncing ball type periodic orbit of Fig.~\ref{fig1} (c) consists of the bouncing ball type periodic orbit in the individual disks and the shortest path between them which is the only periodic orbit outside the two disks. As these periodic orbits [red lines of Fig.~\ref{fig1} (b) and (c)] are made by combining periodic orbits which exist in disks with a closed boundary condition, they always exist in the coupled dielectric disk system, independent of system parameters. The hexagonal-shaped periodic orbit of Fig.~\ref{fig1} (d) is a new periodic orbit for which no corresponding periodic orbits exist in a closed boundary condition. Consequently, its existence and (geometric) properties will depend on the system parameters.

\begin{figure}
\begin{center}
\includegraphics[width=\figsize\textwidth]{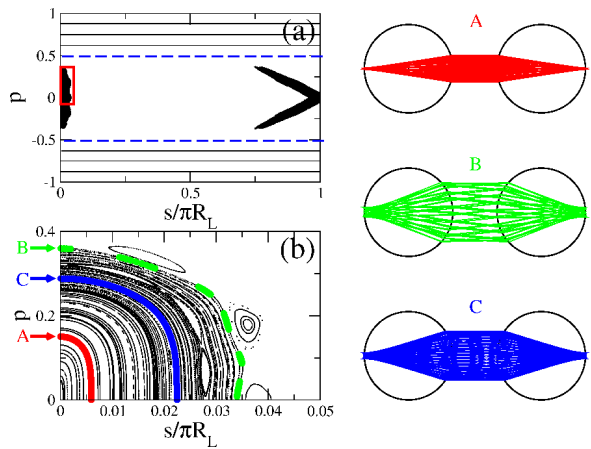}
\caption{(color online) 
(a) The ray attractors in the left disk of coupled dielectric disks with interdisk distance $d=1.0$, ratio of radii $R = R_L / R_R =1.0$, and refractive indices $n=n_L /n_0 = n_R /n_0 =2.0$. The blue dashed line marks the critical line $p_c=1/n=0.5$ for total internal reflection. Note that the right and left part of the attractor are related by the ray dynamics of the ORM within the left disk. 
(b) Enlarged ray attractor corresponding to the red square of (a). The arrows A (lower red), B (upper green), and C (middle blue) mark the attractors that correspond to the red, green, and blue trajectories, respectively, shown on the right.}
\label{fig2}
\end{center}
\end{figure}

\begin{figure}
\begin{center}
\includegraphics[width=\figsize\textwidth]{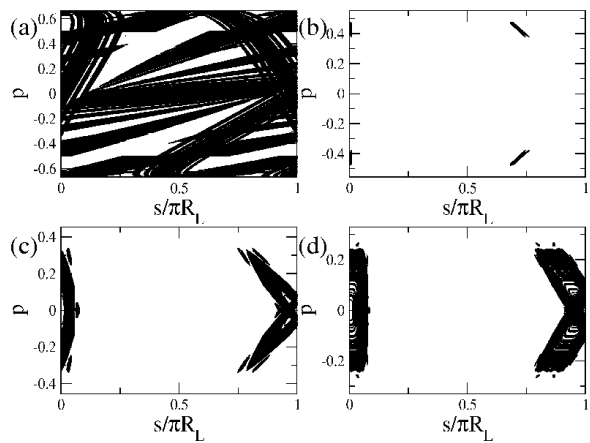}
\caption{PSOSs in the open region ($|p|<p_c$) on the phase space when (a) $n=1.5$, (b) $n=1.8$, (c) $n=2.2$, and (d) $n=3.0$.}
\label{fig3}
\end{center}
\end{figure}

We now study the ray dynamics in our RMDS-model in terms of the Poincar\'e surface of section (PSOS) that is obtained by plotting the position $s$ (arclength along disk boundary, cf.~Fig.~\ref{fig1} (a)) \cite{com0} and the corresponding $p$, sine of the angle of incidence, for each reflection point at the boundary of the left disk. The PSOS for the symmetric system ($n_L/n_0=n_R/n_0 \equiv n$ and $R_L=R_R$, all lengths are measured in units of the interdisk spacing $d \equiv 1.0$) is shown in Fig.~\ref{fig2}. It is obtained from initial rays uniformly distributed in phase space and the ray trajectories resulting from the RMDS for coupled dielectric disks. In the region of total internal reflection ($|p| > p_c$), the ray dynamics follows horizontal lines that are single-disk WGMs. Differences from the single-disk behavior occur, as expected, in the open region ($|p| < p_c$) of phase space where the dielectric coupling between the two disks becomes effective. We emphasize that the ray dynamics will go through a transient behavior before the structure displayed in Fig.~\ref{fig2} is reached in the stationary regime. Since the ray dynamics within the RMDS-model depends on the system parameters (geometrical details, refractive index ratios), so will the details of the PSOS in this region, see also Fig.~\ref{fig3} and the discussion below. 
For the symmetric geometry, the PSOS structure in the open region forms an attractor in the stationary regime, see Fig.~\ref{fig2} (a), i.e., the rays with initial $|p|$ smaller than $p_c$ go to the attractor after a transient time. The black areas in Fig.~\ref{fig2} (a) represent the attractor which actually has regular island structures as shown in Fig.~\ref{fig2} (b). For instance, rays started at positions $(s,p) =$  (0.0,0.15) and (0.0,0.36) on the attractor form the red (A) and green (B) trajectories in configuration space (upper and central panel on the right of Fig.~\ref{fig2}) and the corresponding sets in phase space (marked by arrows). A ray with initial position $(s,p)=(0.0,0.45)$ enters the blue (C) set (lower panel on the right of Fig.~\ref{fig2}) after the ray travels the open region for the transient time and then stays on the attractor forever. 

Figure \ref{fig3} shows the attractor structure in the PSOS for different refractive indices and the same symmetric geometry. Clearly, the details of the attractor depend crucially on the underlying ray dynamics, that is, (besides a geometry dependence) the index $n$ of refraction. For $n=1.5$,  cf.~Fig.~\ref{fig3}(a), the attractor covers a broad range of phase space and its fractal structure represents the fully chaotic ray dynamics. The onset of island formation can be seen for $n=1.8$ in panel (b). Further increasing $n$ enlarges the islands and adds to the richness of their structure, recall the detail shown in Fig.~\ref{fig2}(b) for $n=2$. This behavior is reminiscent of the evolution of KAM islands in the context of chaotic dynamics, e.g., as an external parameter such as the kicking strength in the kicked rotator is changed. In the present case, it is in particular the $n$-dependence of the selection rule for the RMDS that effectively changes the underlying (RMDS) map. The detailed description of the attractors will be reported elsewhere. 

The attracting behavior of the RMDS is caused by the underlying deterministic selection rule, as illustrated in Fig.~\ref{fig3}: The attractor structure depends on the parameter $n$ that in turn determines the selection rule. The basin of attraction is formed by the open region in phase space that extends between momenta $p = \pm 1/n$ (rays started outside this region will remain above/below the critical line and inside the disk they started in forever). When a ray reaches the attractor after some transition time, it then has to stay on the attractor forever, according to the rules of the RMDS: Rays that would leave the attractor are never selected. Note that this also implies that no attractors are formed if the full ray dynamics, combining the features of the RMDS and the ordinary ray model (ORM) of the individual disks, is taken into account. Therefore, the attractors indicate invariant sets to which ray trajectories approach in the dynamics determined by the RMDS. Notice that these attractors are similar to the ``quasi-attractors'' in a piecewise smooth area-preserving map for which noninvertibility was found to induce the phase space collapse \cite{Wan01_1,Wan01_2}. We also point out that the attractors that we study here are different from those described in Ref.~\cite{Alt08} which resulted in the context of optical microcavities with a corrected ray dynamics.

A striking feature of the RMDS attractors is the existence of regular islands. For instance, although the periodic orbit of Fig.~\ref{fig1} (c) consists of marginally stable periodic orbits inside the disks and an unstable periodic orbit connecting the disks, the combined RMDS periodic orbit is marginally stable with elliptic structure as shown in Fig.~\ref{fig2} (b). This is confirmed when studying its stability via its monodromy matrix $\mathbf{M}$ that, in terms of the Birkhoff coordinates, is defined by \cite{Ber91}
\begin{equation}
\left( \begin{array}{c}
\delta s_j \\
\delta p_j \\
\end{array} \right)=
\mathbf{M}
\left( \begin{array}{c}
\delta s_{j-1} \\
\delta p_{j-1} \\
\end{array} \right)=
\left( \begin{array}{cc}
m_{11} & m_{12} \\
m_{21} & m_{22} \\
\end{array} \right)
\left( \begin{array}{c}
\delta s_{j-1} \\
\delta p_{j-1} \\
\end{array} \right),
\label{eq1}
\end{equation}
where $s_{j}$ is the arclength $s$ at the $j$th bounce point and angular momentum $p_{j}$ is $p$ at the $j$th bounce point. A trajectory starts from a position $s_{j-1}$ with momentum $p_{j-1}$ and ends at a position $s_{j}$ with momentum $p_{j}$. We then take another trajectory starting with slightly different initial conditions $s_{j-1} + \delta s_{j-1}$ and $p_{j-1} + \delta p_{j-1}$. The deviations at the end position are obtained from the deviations at the starting position in linear approximation as Eq.~(\ref{eq1}) \cite{Lic92,Sto99}.

To obtain the monodromy matrix of the periodic orbit of Fig.~\ref{fig1} (c), we divide this orbit into three elements, namely the trajectories inside and in between the disks, and the refraction at the dielectric interfaces \cite{Sie90,Sie98,Pri00}. First, we obtain the monodromy matrix of the trajectory of the periodic orbit inside a disk. The monodromy matrix for a left (right) disk is thus
\begin{equation}
\mathbf{M_{L(R)}}=
\left( \begin{array}{cc}
1 & -2 R_{L(R)} \\
0 & 1 \\
\end{array} \right).
\end{equation}
From the relation $\mathrm{Tr}~\mathbf{M_{L(R)}} = 2$, we confirm that the trajectory inside the disks are marginally stable. Secondly, we obtain the monodromy matrix of the trajectory in between 
the two disks. After some algebra, the corresponding monodromy matrix reads 
\begin{equation}
\mathbf{M_I}=
-\left( \begin{array}{cc}
1+{{d}\over{R_L}} & d \\
{{1}\over{R_L}}+{{1}\over{R_R}}+{{d}\over{R_L R_R}} & 1+{{d}\over{R_R}} \\
\end{array} \right).
\end{equation}
Since $\mathrm{Tr}~\mathbf{M_I} < -2$ if $d>0$, this part of the trajectory is always unstable. Note that up to this point the problem is equivalent to the corresponding orbit in a two-disk billiard \cite{Jos92}. In order to obtain the full monodromy matrix for our coupled-dielectric-disks problem, we need in addition the monodromy matrices at the dielectric interfaces where the light is refracted. Since, according to Snell's law, $\delta p_{j} = n \delta p_{j-1}$ if a ray goes from inside to outside a disk and $\delta p_{j} = {{1}\over{n}} \delta p_{j-1}$ in the opposite case, the respective monodromy matrices are
\begin{equation}
\mathbf{M_{B_1}}=
\left( \begin{array}{cc}
1 & 0 \\
0 & n \\
\end{array} \right)
\end{equation}
and
\begin{equation}
\mathbf{M_{B_2}}=
\left( \begin{array}{cc}
1 & 0 \\
0 & {{1}\over{n}} \\
\end{array} \right).
\end{equation}
Finally, the monodromy matrix of the horizontal bouncing ball type periodic orbit of Fig.~\ref{fig1} (c) is
\begin{equation}
\mathbf{M}=\mathbf{M_L}\mathbf{M_{B_2}}\mathbf{M_I}\mathbf{M_{B_1}}\mathbf{M_R}\mathbf{M_R}\mathbf{M_{B_2}}\mathbf{M_I}\mathbf{M_{B_1}}\mathbf{M_L}.
\end{equation}
Although $\mathbf{M_{B_1}}$ and $\mathbf{M_{B_2}}$ induce area-expanding and area-contracting properties, respectively, upon transmission of rays through dielectric interfaces, the final linear mapping for the full periodic orbit satisfies the area-preserving property with $\det \mathbf{M} = 1$. 

\begin{figure}
\begin{center}
\includegraphics[width=\figsize\textwidth]{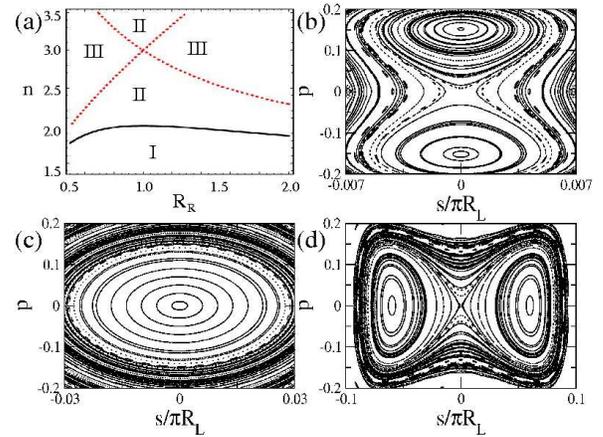}
\caption{(color online) (a) Contour plots which satisfy the conditions of $\mathrm{Tr}~\mathbf{M} = 2$ (black line) and $\mathrm{Tr}~\mathbf{M} = -2$ (red dashed lines) on $(R_{R}, n)$-parameter space when $d=1.0$. PSOSs at (b) $n=1.98$ (region I), (c) $n=2.5$ (region II), and (d) $n=3.0$ (region III) when $d=1.0$ and $R_{R}=1.2$.}
\label{fig4}
\end{center}
\end{figure}

The stability of the horizontal periodic orbit depends on the value of $\mathrm{Tr}~\mathbf{M}$. In the case of $\left|\mathrm{Tr}~\mathbf{M}\right| > 2$, the periodic orbit is linearly unstable but in the case of $\left|\mathrm{Tr}~\mathbf{M}\right| < 2$, the periodic orbit is linearly stable. In order to classify the linear stability of this orbit, contour plots satisfying the condition $\mathrm{Tr}~\mathbf{M} = \pm 2$ in $(R_{R}, n)$-parameter space (with $d=1.0$) are shown in Fig.~\ref{fig4} (a).  The linear stability can be classified into three different types \cite{Lic92,Sto99}.

In region I with $\mathrm{Tr}~\mathbf{M} > 2$, the two eigenvalues of $\mathbf{M}$ are a pair of  reciprocal positive real values and the regular (ordinary) hyperbolic periodic orbit is linearly unstable.
In region II with $-2 < \mathrm{Tr}~\mathbf{M} < 2$, the two eigenvalues of $\mathbf{M}$ form a complex conjugate pair on the unit circle and the corresponding elliptic periodic orbit is linearly stable.
In region III with $\mathrm{Tr}~\mathbf{M} < -2$, the two eigenvalues of $\mathbf{M}$ are a pair of reciprocal negative real values and the inverse (reflection) hyperbolic periodic orbit is linearly unstable. The existence of the periodic orbit, $(s,p)=(0,0)$, is independent of the system parameters but the stability of the periodic orbit is determined by the system parameters in coupled dielectric disks. 

To confirm the stability diagram of Fig.~\ref{fig4} (a), we obtained numerically the PSOSs in the close vicinity of the periodic orbit, more precisely around its point $(s,p)=(0,0)$ at the left disk boundary.  Figure~\ref{fig4} (b), (c), and (d) show the PSOSs in the different stability regions of Fig.~\ref{fig4} (a). We start in region I with a suitable parameter $n=1.98$ such that the horizontal periodic orbit possesses a hyperbolic structure. As $n$ decreases, the separatrix is broken and then the island structure near $(0,\pm 0.15)$ become smaller and finally disappear. In region II, the periodic orbit has elliptic character. In region III, the periodic orbit has again a hyperbolic structure. As $R_R$ increases, the separatrix is also broken and then the island structure near $(\pm 0.05, 0)$ become smaller and finally disappear.

In contrast to hard-wall billiard systems, the decay of the ray intensity in open optical system is an additional important characteristics. When the ray hits the dielectric boundary, the ray intensity decays according to Fresnel's equation \cite{Haw95}. The decay rate $\gamma$ is defined by
\begin{equation}
e^{-\gamma t} = I(t),
\label{eq_ray}
\end{equation}
where $I(t)$ is the remaining ray intensity after time $t$ which is also a measure of the trajectory length. In order to obtain the decay rate $\gamma$ of attractor in RMDS, we consider an ensemble of initial points distributed uniformly over the open region of phase space. After the transient time, we calculate the remaining ray intensity according to Fresnel's equation whenever the discarded ray in RMDS emits from the system. The time $t$ is scaled to be the length of ray trajectory inside two disks and $1/n$ of the length of ray trajectory outside disks. The decay rates of regular islands of Fig.~\ref{fig2} are found to be larger than $0.46$ and smaller than $0.52$ from numerical calculations. In consistency with these values we obtain the decay rate $\gamma_{HBB}$ of the horizontal bouncing ball type periodic orbit as $0.4883$ from Eq.~(\ref{eq_ray}).

Up to now we have studied the ray dynamics of two coupled dielectric disks employing a ray model with a specially chosen selection rule. We have found the formation of attractors as an essential feature. The question arises to what extent the RMDS model accurately describes the reality. To this end we now study the wave dynamics of the system and, following the concept of ray-wave correspondence, will pay special attention to the possibility of modes that correspond to the attractor structures of the ray model description.

\section{Ray-wave correspondence in coupled optical microdisks}

\begin{figure}
\begin{center}
\includegraphics[width=\figsize\textwidth]{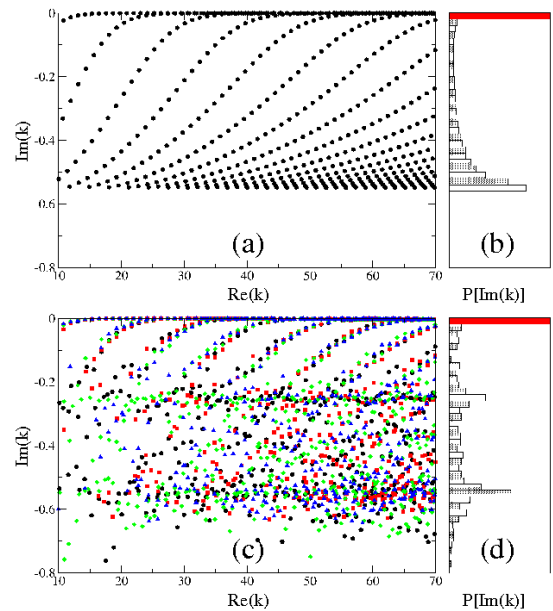}
\caption{(color online) Resonances in (a) single and (c) coupled optical microdisks when $n=2.0$ and $d=1.0$ on complex plane. In (c), the black circles, red rectangles, green diamonds, and blue triangles are resonances of which symmetric classes are EE, OE, EO, and OO, respectively. The distributions $\mathrm{P[Im}(k)]$ of the imaginary parts of complex resonances (b) in a single optical microdisk and (d) with EE-parity in coupled optical microdisks. The red (dark) bars represent high-$Q$ resonances.}
\label{fig5}
\end{center}
\end{figure}

Our next interest is thus 'How do the classical structures emerging from RMDS in coupled optical microdisks relate to resonance modes?' In a single optical microdisk, the properties of the imaginary parts of the complex resonances are well explained by semiclassical analysis \cite{Ryu08}. Resonance modes can be classified into two groups according to their loss which is represented by the imaginary part. Modes of the first group are high-$Q$ WGMs with dominant angular momenta above the critical $p_c$ for total internal reflection and, consequently, only tunneling leakage to the exterior. The second group consists of low-Q modes with dominant angular momenta below $p_c$ implying refractive leakage. Figure~\ref{fig5} (a) and (b) shows resonances with transverse magnetic (TM) polarization of a single disk with $n=2$  in the complex plane that are obtained from the Helmholtz equation \cite{Jac75}, and the corresponding distribution of their imaginary parts. The complex wave number $k$ inside the microdisk is given as $nk_{0}R_R$, where $k_{0}$ is the wave number outside. In this case, the resonances near $\mathrm{Im}(k) \sim 0$ are very high-$Q$ resonances and form the first group indicated by the long red bar in Fig.~\ref{fig5} (b). The resonances near $\mathrm{Im}(k) \sim -0.55$ are almost bouncing ball type resonances with very low-Q factor. The distribution of these low-Q resonances representing the second group as well as the lower limit of imaginary values can be explained by semiclassical analysis.

For the coupled optical microdisk system, we obtained resonances belonging to four symmetry classes (EE, OE, EO, and OO with even (E) and odd (O) symmetries with respect to the vertical and horizontal symmetry axes of the two-disk system) using the boundary element method \cite{Wie03}. They are shown in Fig.~\ref{fig5} (c) for the parameters $d=1.0$, $R=R_L /R_R =1.0$, and $n=n_L /n_0 = n_R / n_0 = 2.0$. The distribution of imaginary parts of the resonances with EE-parity is presented in Fig.~\ref{fig5} (d) \cite{com2}. In contrast to the two peak structure found for the single disk, there are now three prominent peaks in the distribution of imaginary parts of resonances in coupled optical microdisks.

First, the very high-$Q$ WGMs with imaginary parts larger than $-0.02$ are almost the same as those in a single optical microdisk because the interdisk distance which is inversely proportional to coupling strength is sufficiently large. For the high-$Q$ WGMs, the two-disk system can be considered as consisting of two individual microdisks and a coupling term. The coupling between two microdisks can be explained by optical tunneling or frustrated total internal reflection \cite{Zhu86,Cha05}. The tunneling process occurs at the interdisk spacing which is a small part of the circular boundary and does not significantly influence the resonance position.
Also, the high-$Q$ resonances with imaginary parts between -0.2 and -0.02 do not differ a lot from those in a single optical microdisk but the resonances split according to the symmetry classes and these splittings tend to increase as the imaginary parts become smaller. It is natural that the coupling strength, represented by the amount of splitting, is proportional to the resonance loss because larger emission from one of the microdisks induces a larger coupling to the other microdisk. 

Secondly, the peak in the distribution near $\mathrm{Im}(k) \sim -0.55$ corresponds to the bouncing ball type resonances in a single optical microdisk and is therefore directly related to the single disk result in Fig.~\ref{fig5} (b). Although the results presented are for the TM case, there is generally no lower limit of imaginary values because the mode coupling to the environment is not intrinsically limited. Rather, the coupling between internal and external (or outer or shape \cite{Dub08,Bog08,Det09,Cho10}) resonances, i.e. modes localized inside and outside the cavity respectively, causes as broad distribution of imaginary parts.

\begin{figure}
\begin{center}
\includegraphics[width=\figsizesmall\textwidth]{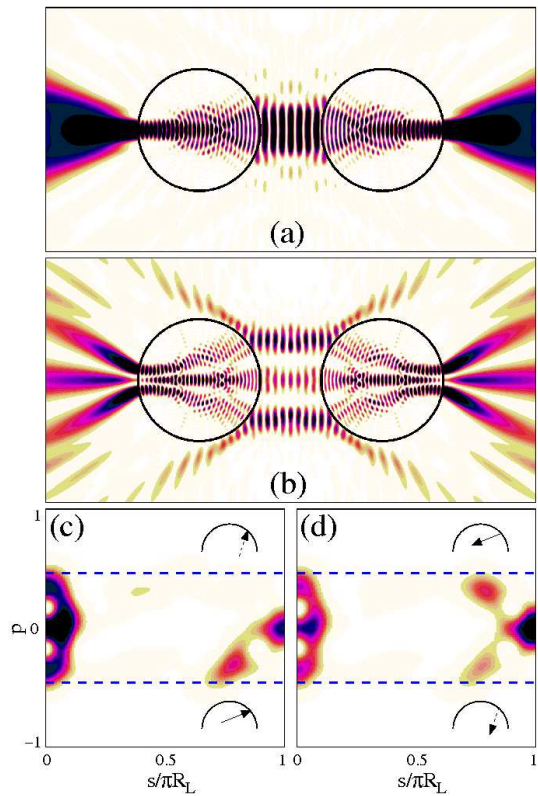}
\caption{(color online) Near field intensity patterns of resonance modes localized on regular islands with wave numbers (a) $k=45.0025 - i 0.2451$ and (b) $k=50.9680 - i 0.2582$. Black-red(gray)-yellow(light gray)-white colors indicate high to low intensity. (c) Incident and (d) emerging generalized Husimi functions taken at the inner boundary of the left disk, the detail shown corresponds to the PSOS representation in Fig.~\ref{fig2} (a). The dashed lines $|p| = p_c=1/n=0.5$ mark the critical momemtum for the onset of total internal reflection. The arrows of insets are the ray trajectories corresponding to positions, $(s,p) \sim (0.8,\pm0.3)$, of upper and lower lobes of the right islands.}
\label{fig6}
\end{center}
\end{figure}

The striking contrast between the distributions in single and coupled optical microdisks is the peak near $\mathrm{Im}(k) \sim -0.25$ in Fig.~\ref{fig5} (d). Those resonances do not have a partner in the single optical microdisk. In fact, these resonances originate from the new classical structure obtained within the RMDS ray model, namely the attractors discussed above. This is confirmed in Fig.~\ref{fig6} that shows the near field intensity patterns of two typical regular modes corresponding to the island structure (RMCI) of the RMDS attractors in Fig.~\ref{fig2}. The patterns of Fig.~\ref{fig6} (a) and (b) resemble the horizontal bouncing-ball-type periodic orbit in Fig.~\ref{fig1} (c) and the ray trajectories of the RMDS in Fig.~\ref{fig2}, respectively. Figure \ref{fig6} (c) and (d) represent the generalized Husimi functions \cite{Hen03} of the mode in Fig.~\ref{fig6} (b) and show incident and emerging intensities, respectively, at the inner boundary of upper semicircle of the left disk. The detail shown can be directly compared to the PSOS shown in Fig.~\ref{fig2} (a), and we find indeed a convincing correspondence to the island structure shown there. The Husimi functions follow it very closely, except for the missing of the upper part of the right island in panel (c), indicating that there is (almost) no incident intensity coming from inside the left disk, that is actually accompanied by a difference in intensity in the upper and lower lobes of this island in panel (d) for the emerging Husimi function. As the RMDS-based PSOS does, unlike the Husimi functions, not distinguish between incoming and outgoing rays it is in fact the sum of two Husimi functions that has to be compared to the PSOS island structure in Fig.~\ref{fig2} (a), resulting in a nice agreement. The ray trajectories of insets in Fig.~\ref{fig6} (c) and (d) show the correspondence between the near field pattern of Fig.~\ref{fig6} (b) and the Husimi functions. The solid- and dashed-line arrows represent high and low intensities, respectively. The convincing agreement found in phase and real space, and especially the clear peak near $\mathrm{Im}(k) \sim -0.25$ in Fig.~\ref{fig5} (d) suggest that indeed many resonances are closely associated with the RMDS island structure, which justifies in turn the use and the validity of the RMDS that we introduced and applied in the previous Section.

In order to elucidate the correspondence between the decay rate $\gamma$ of the islands and the loss of resonances, we introduce the intensity of resonance decay given by
\begin{equation}
I(t)=\left|e^{-i \omega t}\right|^2 = \left|e^{-i (\mathrm{Re}(k) + i \mathrm{Im}(k)) t}\right|^2 = e^{2 \mathrm{Im}(k) t},
\label{eq_res}
\end{equation}
where we take the speed of light $c=1$ in vacuum without loss of generality. From Eq.~(\ref{eq_ray}) and Eq.~(\ref{eq_res}), we obtain the relation
\begin{equation}
\mathrm{Im}(k) = - \gamma / 2.
\label{eq10}
\end{equation}
Since the island structure has decay rates $\gamma$ between $0.46$ and $0.52$ as we obtained in previous section, RMCIs have imaginary values between $-0.23$ and $-0.26$, in full agreement with the numerical result $\mathrm{Im}(k) \sim -0.25$ read off from Fig.~\ref{fig5} (d). We conclude that the new classical structures found within the RMDS, such as the regular islands of Fig.~\ref{fig2}, play an important role and can host RMCIs. As a consequence, the distribution of the resonance losses is changed with respect to the uncoupled system. Moreover, the RMDS explains well the losses of RMCIs as well as the near field intensity patterns.

\begin{figure}
\begin{center}
\includegraphics[width=\figsizesmall\textwidth]{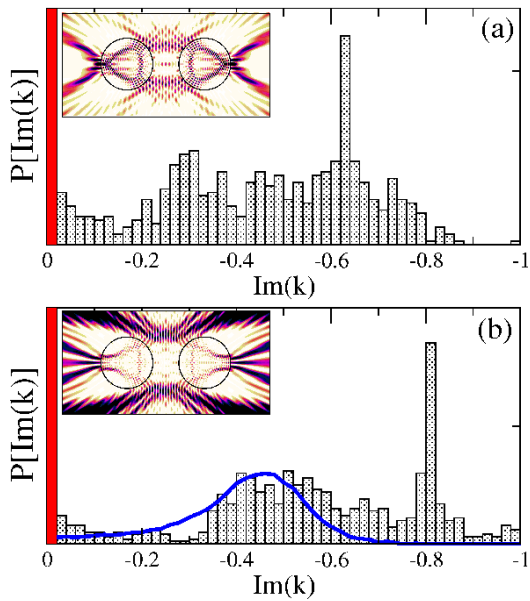}
\caption{(color online) The distributions of the imaginary parts of resonances in coupled optical microdisks when (a) $n=1.8$ and (b) $n=1.5$ with $d=1.0$. Insets are near field intensity patterns of typical modes corresponding to the classical structure of RMDS, of which $k=45.7196 - i 0.2084$ and $k=47.9012 - i 0.5649$, respectively. The blue line is the distribution numerically obtained from the decay rates of the underlying classical strucuture of Fig.~\ref{fig3} (a) and Eq.~(\ref{eq10}).}
\label{fig7}
\end{center}
\end{figure}

Figure~\ref{fig7} (a) and (b) show the distributions of imaginary parts of complex resonances when $n=1.8$ and $n=1.5$, respectively. The distributions have also a three peak structure, but the central peak is less pronounced and broader than in the case of $n=2.0$. The broad distributions originates from the chaotic-to-regular transition of classical attractor island structure when the refractive index $n$ is increased from 1.5 to 1.8, cf.~Fig.~\ref{fig3}. Below $n \sim 1.732$, the PSOS in the open region ($|p|<p_c$) is fully chaotic. When $n=1.8$, only small regular islands in the vicinity of the hexagonal shaped periodic orbit of Fig.~\ref{fig1} (d) appear in the PSOS. The inset of Fig.~\ref{fig7} (a) shows a corresponding RMCI ($\mathrm{Im}(k)=-0.2084$, this value is consistent with the decay rate $\gamma \sim 0.4$ of the hexagonal shaped periodic orbit). For $n=1.5$, the near field intensity pattern shown in the inset of Fig.~\ref{fig7} (b) is typical for a chaotic mode with a rather low $\mathrm{Im}(k)=-0.5649$ and correspondingly high intensity outside the two microdisks. The distribution of the imaginary parts of resonances in a fully chaotic system is associated with the decay rate of the underlying classical structure of Fig.~\ref{fig3} (a), which is represented by the blue line in Fig.~\ref{fig7} (b). Recently, it has been investigated that this distribution can also be related to the escape rate for the phase space points near the chaotic repeller including Fresnel's laws \cite{Wie08_2,Sch09}.

\begin{figure}
\begin{center}
\includegraphics[width=0.45\textwidth]{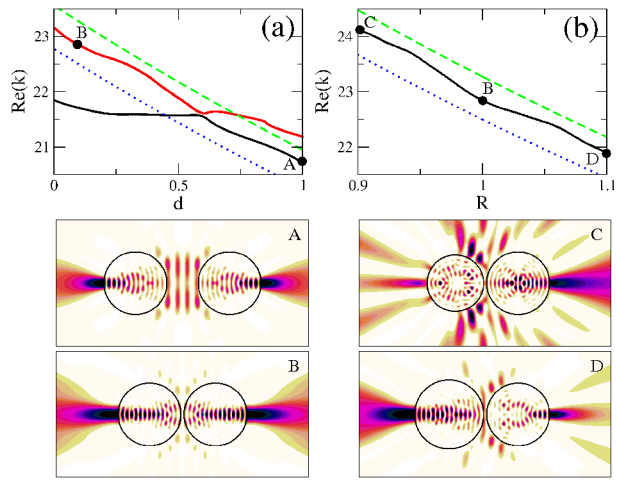}
\caption{(color online) The real parts of resonances as a function of (a) the interdisk distance in coupled identical optical microdisks and (b) the ratio of radii in coupled nonidentical optical microdisks when $d=0.1$. The four near field intensity patterns are for A-, B-, C-, and D-modes which are represented by black circles in (a) and (b).}
\label{fig8}
\end{center}
\end{figure}

We now consider the dependence of the resonances and the loss distribution on the system parameters. For the case of the high-$Q$ WGMs, the real parts of the resonances split as the interdisk distance decreases if the two microdisks are sufficiently close \cite{Ryu06} and is inversely proportional to the radius of a microdisk or independent of the radius in coupled nonidentical optical microdisks because the dominant WGM locates on only one microdisk \cite{Ryu09}. Concerning the horizontal bouncing ball type periodic orbit of Fig.~\ref{fig1},  Fig.~\ref{fig8} (a) and (b) show the real parts of the corresponding RMCIs as function of the interdisk distance and the ratio of radii $R$ of the two microdisks, respectively. Unlike the real parts of high-$Q$ resonances, the real parts of RMCIs increase as the interdisk distance or/and the radius of the left microdisk decreases as shown in Fig.~\ref{fig8} (a) and (b), respectively. The black and red full lines in Fig.~\ref{fig8} (a) represent RMCIs which localize on the horizontal periodic orbit. They undergo an avoided resonance crossing near $d=0.6$: For lower $d$, the resonance is of type B and follows the red line; after the avoided crossing, it follows the black line and changes its character to A-type. As the near field intensity patterns of the A- and B-modes in Fig.~\ref{fig8} show, both modes have the mode index $l=15$ on the periodic orbit. The resonances depending on the radius of the left microdisk in Fig.~\ref{fig8} (b) are also RMCIs with localized intensity patterns on the periodic orbit; their near field intensity patterns are shown in Fig.~\ref{fig8}.

Using the quantization rule on the periodic orbit, we obtain the real part $\mathrm{Re}(k)$ of a resonance as a function of the radii, $R_{L}$ and $R_{R}$, of the microdisks, the interdisk distance $d$, and assuming closed boundary conditions at both ends of the periodic orbit, as 
\begin{equation}
k=\frac{\alpha \pi}{2(R_{L}+R_{R})+d/n},
\label{quantize}
\end{equation}
with $\alpha = 2l$ and $2l-1$ for Dirichlet and Neumann boundary conditions, respectively. For TM polarization, the $\mathrm{Re}(k)$ satisfies the relation
\begin{equation}
k_{N} < \mathrm{Re}(k) < k_{D},
\label{qu}
\end{equation}
where $k_{N}$ and $k_{D}$ are the wave numbers for Neumann and Dirichlet boundary conditions, respectively. In Fig.~\ref{fig8}, the green dashed and blue dotted lines represent $k_D$ and $k_N$, respectively, illustrating that the relation Eq.~(\ref{qu}) is well satisfied.

We finally mention that coupling-originated classical structures such as the attractor islands discussed here may also affect the near field pattern of high-$Q$ modes. Whereas it is generally accepted \cite{Sch04,Lee05,Shi06,Lee207,Wie08} that the low-intensity tail structure of the near field patterns of high-$Q$ modes in chaotic microcavities are related to the chaotic repeller structure which corresponds to the unstable manifold structure near the critical line, this might change when new regular structures appear in the phase space due to the presence of a second disk to which the light can be coupled. Then, a hybridization \cite{Wie06} of high-$Q$ WGM (single disk) and low-Q RMCI (coupled disks) is possible and was confirmed in numerical calculations. In this case, the near field pattern of the high-$Q$ mode is not determined by the chaotic repeller structure near the critical lines, but rather by elements of the low-Q mode leakage which are typically positioned in the center of the leaky region instead.

\section{Summary}

We have proposed the ray model with deterministic selection rule (RMDS) for coupled optical microdisks as a theoretical model to deal with the inherent ray splitting dynamics in optical systems. We found that new classical structures in the form of regular islands (representing the attractor of the RMDS model) occurred, for a certain range of system parameters, in phase space. We have elucidated the stabilities and the decay rates of these structures which depend on the system parameters such as interdisk distance, ratio of radii, and refractive indices of microdisks. 

We have confirmed the physical significance of structures emerging from the RMDS by identifying resonances associated with the classical structures and investigating the ray-wave correspondence. The near field intensity patterns of the resonances resemble the trajectories of regular islands and the real and imaginary parts can be explained by quantization rules and decay rates of ray trajectories, respectively. We expect that the ray model with a specific deterministic selection rule will be useful for the description of resonance modes in other coupled systems and generally in complex physical systems with splitting dynamics.

\section*{Acknowledgment}

We thank H. Kantz, J.-B. Shim, S. Shinohara, J. Unterhinninghofen, and J. Wiersig for discussions. We gratefully acknowledge financial support by the German Research Foundation (DFG)  within the Research Unit FG 760 and the Emmy-Noether Programme (M.H.).


\begin{thebibliography}{150}
\bibitem{com0} The coordinate $s$ is the arc length of upper semicircle of left disk. The $s$ is sufficient to explore our system because of the symmetries of coupled identical dielectric disks. For convenience, we sometimes use the lower semicircle and the $s$ is a negative value in that case.
\bibitem{com2} We show only the distribution of EE-parity because four distributions of imaginary parts of resonances with EE-, EO-, OE-, and OO-parities are similar.
\end{thebibliography}
\end{document}